\documentclass[twocolumn,prl]{revtex4}

\usepackage[T1]{fontenc}
\usepackage{amsmath,amsbsy,amssymb,graphicx}
\usepackage{times}

\def\Name#1{#1,}
\def\Book#1{\textit{#1}}
\def\Review#1{#1}
\def\Publ#1{(#1)}
\def\Vol#1{\textbf{#1}}
\def\Year#1{(#1)}
\def\Page#1{#1}

\begin{document}

\title{{\Large Dirac Theory and Topological Phases of Silicon Nanotube}}
\author{Motohiko Ezawa}
\affiliation{Department of Applied Physics, University of Tokyo, Hongo 7-3-1, 113-8656,
Japan }

\begin{abstract}
Silicon nanotube is constructed by rolling up a silicene, i.e., a monolayer of
silicon atoms forming a two-dimensional honeycomb lattice.
It is a semiconductor or an insulator owing to relatively large spin-orbit interactions induced
by its buckled structure. 
The key observation is that this buckled structure allows us to control
the band structure by applying electric field $E_z$.
When $E_z$ is larger than a certain critical value $E_{\text{cr}}$, 
by analyzing the band structure and also on the basis of the effective Dirac theory,
we demonstate the emergence of four helical zero-energy
modes propagating along nanotube.
Accordingly, a silicon nanotube contains three regions, namely, a
topological insulator, a band insulator and a metallic region separating
these two types of insulators. 
The wave function of each zero mode is localized within the metallic region,
which may be used as a quantum wire to transport spin currents in future spintronics.
We present an analytic expression of the wave function for each helical zero mode.
These results are applicable also to germanium nanotube.
\end{abstract}

\maketitle

\section{Introduction}

\begin{figure}[t]
\centerline{\includegraphics[width=0.49\textwidth]{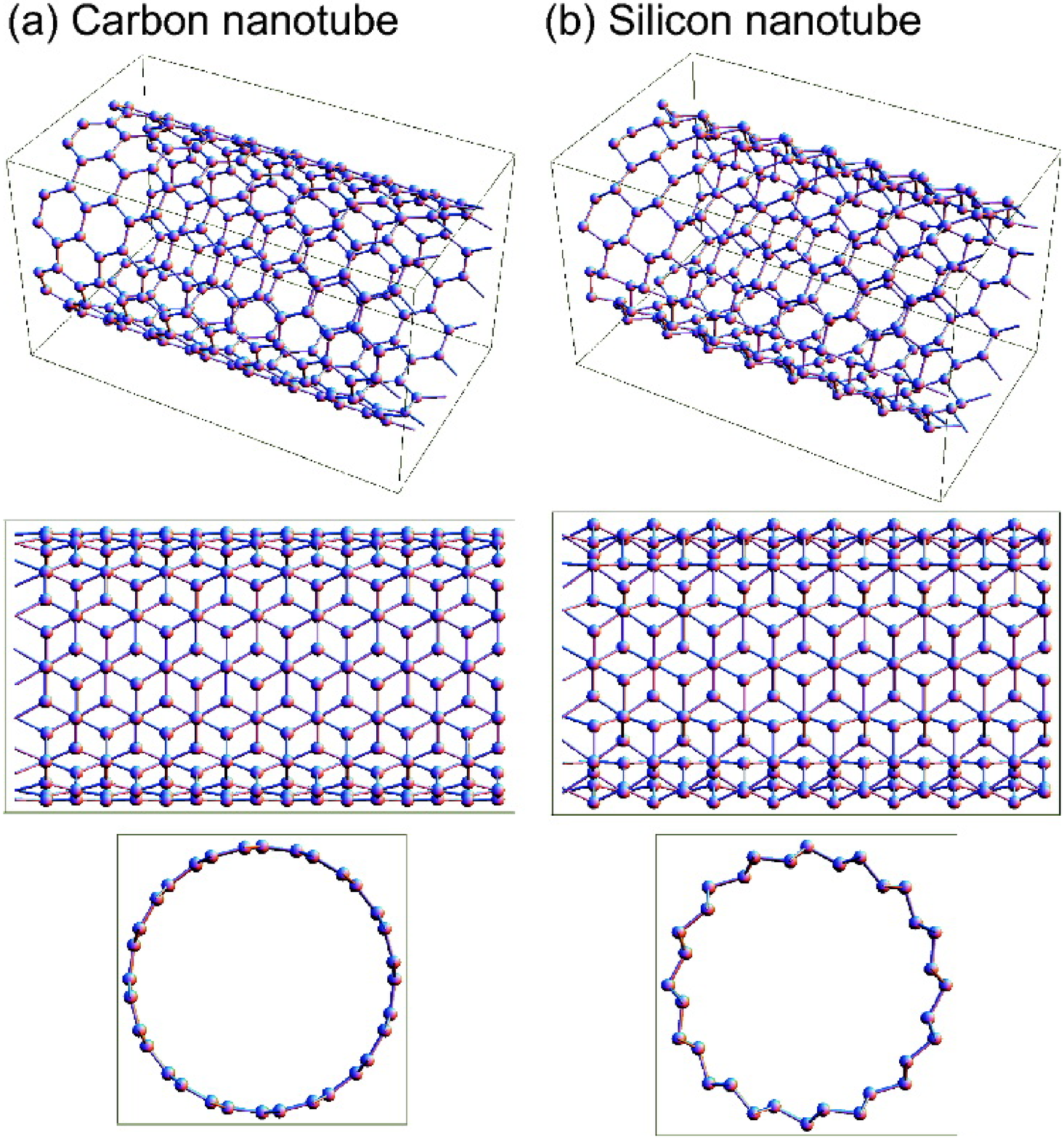}}
\caption{(a) Carbon nanotube and (b) silicon nanotube. The lattice is
distorted due to a large ionic radius of a silicon atom and forms a buckled
structure in silicon nanotube.}
\label{Fig3DTube}
\end{figure}

Carbon nanotube is one of the most fascinating materials. There is a variety
of nanotubes from metal to insulator depending on how it is constructed by
rolling up a graphene\cite{Saito, AndoReview}. Similarly, silicon nanotube
may be constructed by rolling up a silicene\cite%
{Shiraishi,Aufray,LiuPRL,EzawaNJP}, a monolayer of silicon atoms forming a
two-dimensional honeycomb lattice. Silicon nanotubes have already been
manufactured\cite{Sha,Cres, Small}. Almost every striking property of carbon
nanotube is expected to be transferred to this innovative material since
carbon and silicon belong to the same family in the periodic table.
Nevertheless there exists a major difference. A large ionic radius of
silicon induces a buckled structure\cite{LiuPRL}, which results in a
relatively large spin-orbit (SO) gap of $1.55$meV. Accordingly, silicon
nanotube has a finite gap and it is always a semiconductor or an insulator.
One might think it is an ordinary bund insulator just as most carbon
nanotubes are. See Fig.\ref{Fig3DTube}\ for an illustration of carbon
nanotube and silicon nanotube.

In this paper we reveal an amazing property of silicon nanotube thanks to
this buckled structure when it is placed in external electric field.
Analyzing the band structure in the presence of electric field $E_{z}$
perpendicular to the nanotube axis, we demonstrate that silicon nanotube is
actually a topological insulator. When $E_{z}$ is beyond a certain critical
field $E_{\text{cr}}$, we find four zero-energy modes to emerge in the bulk
band gap and form metallic regions along a nanotube. A silicon nanotube is
made of three different phases, the topological insulator region, the bulk
insulator region and the metallic region separating them.

Topological insulator\cite{Hasan,Qi} is a new state of quantum matter
characterized by a full insulating gap in the bulk and gapless edges
topologically protected. These states are made possible due to the
combination of the SO interaction and the time-reversal symmetry. The
two-dimensional topological insulator is a quantum spin Hall (QSH) insulator
with helical gapless edge modes\cite{Wu}, which is a close cousin of the
integer quantum Hall state. QSH insulator was proposed by Kane and Mele in
graphene\cite{KaneMele}. However, since the SO gap is rather weak in
graphene, the QSH effect can occur in graphene only at unrealistically low
temperature\cite{Min, Yao}. Our finding is that it is materialized naturally
in silicon nanotube.

As we have stated, there emerge four zero-energy modes in silicon nanotube
under uniform electric field. They are helical edge modes of a topological
insulator and propagate along a nanotube: They transport only spins without
charges. These observations are supported by the effective Dirac theory. We
construct explicitly the wave functions describing the four helical zero
modes. In conclusion, we are able to realize a dissipationless spin current
along a silicon nanotube by applying uniform electric field. It may be used
as a quantum wire in future spintronics.

\section{Silicene and tight-binding model}

The band structure of a silicon nanotube is obtained simply by imposing a
certain periodic boundary condition to a silicene sheet provided the
diameter is  large enough. Silicene consists of a honeycomb lattice of
silicon atoms with two sublattices made of A sites and B sites. The states
near the Fermi energy are $\pi $ orbitals residing near the K and K' points
at opposite corners of the hexagonal Brillouin zone. We refer to the K or K'
point also as the K$_{\eta }$ point with the valley index $\eta =\pm 1$. Due
to the buckled structure the two sublattice planes are separated by a
distance, which we denote by $2\ell $ with $\ell =0.23$\AA . It generates a
staggered sublattice potential $\varpropto 2\ell E_{z}(x,y)$ between silicon
atoms at A sites and B sites in external electric field $E_{z}(x,y)$.

The silicene system is described by the four-band second-nearest-neighbor
tight binding model\cite{LiuPRB},

\begin{align}
H& =-t\sum_{\left\langle i,j\right\rangle \alpha }c_{i\alpha }^{\dagger
}c_{j\alpha }+i\frac{\lambda _{\text{SO}}}{3\sqrt{3}}\sum_{\left\langle
\!\left\langle i,j\right\rangle \!\right\rangle \alpha \beta }\nu
_{ij}c_{i\alpha }^{\dagger }\sigma _{\alpha \beta }^{z}c_{j\beta }  \notag \\
& -i\frac{2}{3}\lambda _{\text{R}}\sum_{\left\langle \!\left\langle
i,j\right\rangle \!\right\rangle \alpha \beta }\mu _{i}c_{i\alpha }^{\dagger
}\left( \mathbf{\sigma }\times \hat{\mathbf{d}}_{ij}\right) _{\alpha \beta
}^{z}c_{j\beta }  \notag \\
& +\ell \sum_{i\alpha }\mu _{i}E_{z}^{i}c_{i\alpha }^{\dagger }c_{i\alpha },
\label{BasicHamil}
\end{align}%
where $c_{i\alpha }^{\dagger }$ creates an electron with spin polarization $%
\alpha $ at site $i$, and $\left\langle i,j\right\rangle /\left\langle
\!\left\langle i,j\right\rangle \!\right\rangle $ run over all the
nearest/next-nearest neighbor hopping sites. The first term represents the
usual nearest-neighbor hopping with the transfer energy $t=1.6$eV. The
second term represents the effective SO coupling with $\lambda _{\text{SO}%
}=3.9$meV, where $\mathbf{\sigma }=(\sigma _{x},\sigma _{y},\sigma _{z})$ is
the Pauli matrix of spin, $\nu _{ij}=\left( \mathbf{d}_{i}\times \mathbf{d}%
_{j}\right) /\left\vert \mathbf{d}_{i}\times \mathbf{d}_{j}\right\vert $
with $\mathbf{d}_{i}$ and $\mathbf{d}_{j}$ the two bonds connecting the
next-nearest neighbors. The third term represents the Rashba SO coupling
with $\lambda _{\text{R}}=0.7$meV, where $\mu _{i}=\pm 1$ for the A (B)
site, and $\hat{\mathbf{d}}_{ij}=\mathbf{d}_{ij}/\left\vert \mathbf{d}%
_{ij}\right\vert $. The forth term is the staggered sublattice potential
term. The same Hamiltonian as can be used to describe germanene, which is a
honeycomb structure of germanium\cite{LiuPRL,LiuPRB}, where various
parameters are $t=1.3$eV, $\lambda _{\text{SO}}=43$meV, $\lambda _{\text{R}%
}=10.7$meV and $\ell =0.33$\AA .

By diagonalizing the Hamiltonian (\ref{BasicHamil}) under uniform electric
field $E_{z}$, the band gap $\Delta \left( E_{z}\right) $ of silicene is
determined to be%
\begin{equation}
\Delta \left( E_{z}\right) =2\left\vert \ell E_{z}-\eta s_{z}\lambda _{\text{%
SO}}\right\vert  \label{gap}
\end{equation}%
at the K$_{\eta }$ point, where $s_{z}=\pm 1$ is the electron spin. The gap $%
\Delta \left( E_{z}\right) $ closes at $E_{z}=\eta s_{z}E_{c}$ with%
\begin{equation}
E_{\text{cr}}=\lambda _{\text{SO}}/\ell =17\text{meV/\AA },  \label{CritiE}
\end{equation}%
where it is a semimetal due to gapless modes. It has been shown\cite%
{EzawaNJP} that silicene is a topological insulator for $\left\vert
E_{z}\right\vert <E_{\text{cr}}$, while it is a balk insulator for $%
\left\vert E_{z}\right\vert >E_{\text{cr}}$, as illustrated in Fig.\ref%
{FigRibbonBand}. Hence a topological phase transition occurs between a
topological insulator and a band insulator as $E_{z}$ changes.

\begin{figure}[t]
\centerline{\includegraphics[width=0.4\textwidth]{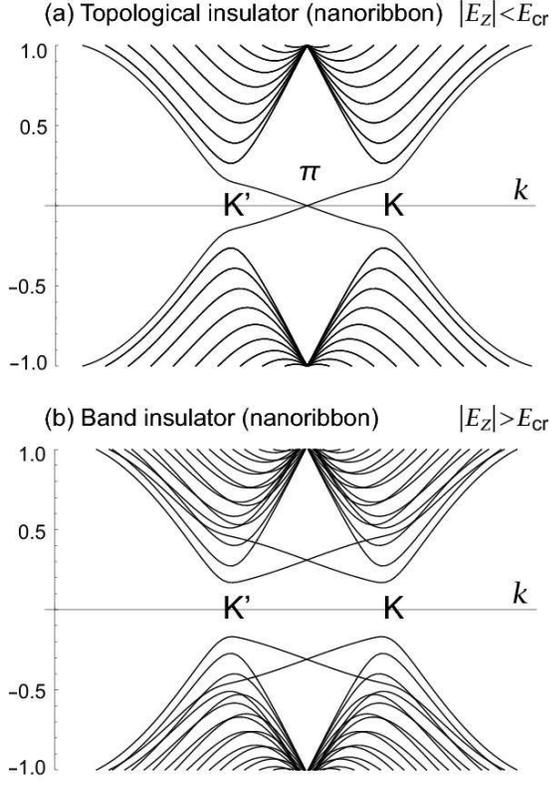}}
\caption{(Color online) (a) The band structure of silicene nanoribbon for $%
|E_{z}|<E_{\text{cr}}$. There appear two bands crossing the gap since there
are two edges, indicating it is topological insulator. When the Rashba
interaction is neglected ($\protect\lambda _{\text{R}}=0$), each band
contains two-fold degenerate zero-energy states corresponding to up and down
spins. (b) The band structure of silicene nanoribbon for $|E_{z}|>E_{\text{cr%
}}$. All states are gapped, and it is bulk insulator. The horizontal axis is
the momentum, and the vertical axis is the energy in unit of the transfer
energy $t$.}
\label{FigRibbonBand}
\end{figure}

The topological insulator is characterized by one of the following two
defining properties\cite{Hasan,Qi}. (1) The topological insulator has a
nontrivial topological number, the $\mathbb{Z}_{2}$ index\cite{KaneMele},
which is defined only for a gapped state. (2) There emerge gapless modes in
the edges [Fig.\ref{FigRibbonBand}(b)]. These two properties are closely
related one to another. Indeed, the reason why gapless modes appear in the
edge of a topological insulator is understood as follows. When a topological
insulator has an edge beyond which the region has the trivial $\mathbb{Z}%
_{2} $ index, the band must close and yield gapless modes in the interface.
Otherwise the $\mathbb{Z}_{2}$ index cannot change its value across the
interface.

The above criteria cannot be applicable to a nanotube as they are since it
is intrinsically one-dimensional (1D). We overcome the problem by applying
electric field and by creating a domain with 1D edges in the surface of a
nanotube: See Fig.\ref{FigHelicTube} we derive later.

\section{Silicon nanotube}

A silicon nanotube can be constructed by rolling up a silicene precisely
just as a carbon nanotube is constructed by rolling up a graphene\cite%
{Saito, AndoReview}. There are a variety of ways how to rolling up a
silicene as in the case of carbon nanotubes. It is specified by the chiral
vector,%
\begin{equation}
\mathbf{L}=n_{1}\mathbf{a}_{1}+n_{2}\mathbf{a}_{2},
\end{equation}%
where $\mathbf{a}_{1}$ and $\mathbf{a}_{2}$ are the basis vectors of the
honeycomb lattice with integers $n_{1}$ and $n_{2}$. Let us take the
coordinate$\ (x,y)$ with the $x$-axis parallel to the chiral vector $\mathbf{%
L}$ (the circumference direction) and the $y$-axis orthogonal to it (the
nanotube axis direction). The nanotube's circumference is given by%
\begin{equation}
L=\left\vert \mathbf{L}\right\vert =a\sqrt{n_{1}^{2}+n_{2}^{2}-n_{1}n_{2}},
\end{equation}%
where $a=3.86$\AA\ is the lattice constant. Let us denote the conjugate
momemtum as $(\hbar k_{x},\hbar k_{y})$. We impose the periodic boundary
condition $\psi (x+L,y)=\psi (x,y)$. It makes the wave vector $k_{x}$
discrete, i.e., 
\begin{equation}
k_{x}=2\pi j/L,\quad \quad j=0,1,\cdots ,2n-1,  \label{CondiKx}
\end{equation}%
where $2n$ is the number of silicon atoms per unit cell. The wave vector $%
k_{y}$ remains continuous within the 1D first Brillouin zone,%
\begin{equation}
-\pi /T\leq k_{y}\leq \pi /T,  \label{CondiKy}
\end{equation}%
where $T$ is the length of the primitive translation vector in the $y$
direction. The 1D energy bands are given by the straight lines specified by (%
\ref{CondiKx}) and (\ref{CondiKy}) in the $(k_{x},k_{y})$ space. The
simplest nanotube is the armchair type, where $L=\sqrt{3}na$.

The energy spectrum of a silicon nanotube is simply given by%
\begin{equation}
E_{j}\left( k_{y}\right) =\mathcal{E}(k_{x},k_{y}),
\end{equation}%
in terms of the energy spectrum $\mathcal{E}(k_{x},k_{y})$ of silicene. The
energy spectrum of a carbon nanotube can be gapless since graphene has a
gapless band. There is a variety of carbon nanotubes from metal to insulator
depending on the way how to choose the chiral vector $\mathbf{L}$. On the
contrary, a silicon nanotube has always a finite gap since silicene has a
finite gap. The band gap is rather insensitive to the choice of $\mathbf{L}$.

\begin{figure}[t]
\centerline{\includegraphics[width=0.4\textwidth]{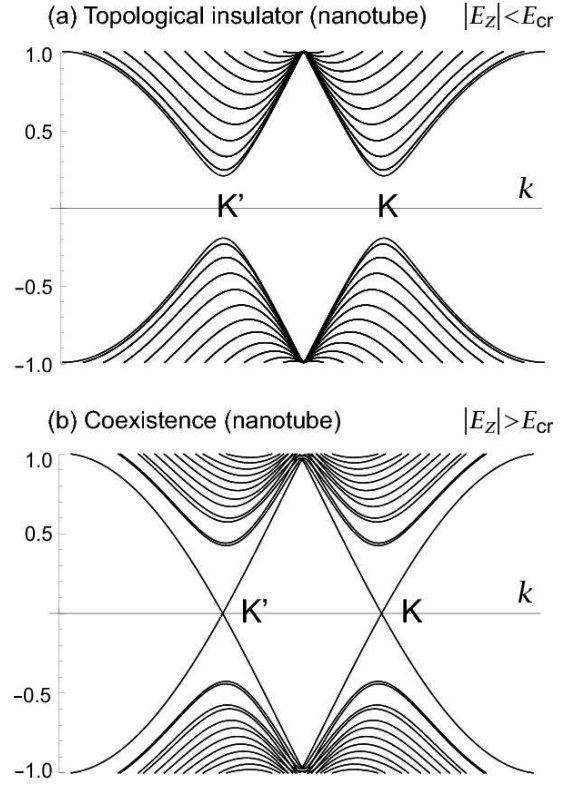}}
\caption{(Color online) (a) The band structure of silicon nanotube for $%
|E_{z}|<E_{\text{cr}}$. All states are gapped, and it is topological
insulator. (b) The band structure of silicon nanotube for $|E_{z}|>E_{\text{%
cr}}$. There appear four bands crossing the gap. They are the zero-energy
states separating the topological and band insulating states in the nanotube
surface: See Fig.\protect\ref{FigHelicTube}. When the Rashba interaction is
neglected ($\protect\lambda _{\text{R}}=0$), each band contains two-fold
degenerate zero-energy states corresponding to up and down spins. The
horizontal axis is the momentum, and the vertical axis is the energy in unit
of the transfer energy $t$.}
\label{FigTubeBand}
\end{figure}

We have calculated the band structure of an armchair silicon nanotube based
on the tight-binding model (\ref{BasicHamil}) by changing the uniform
external electric field $E_{z}$. The band structure is strikingly different
for $\left\vert E_{z}\right\vert <E_{\text{cr}}$ and $\left\vert
E_{z}\right\vert >E_{\text{cr}}$, as illustrated in Fig.\ref{FigTubeBand}.
All states are gapped for $\left\vert E_{z}\right\vert <E_{\text{cr}}$. On
the other hand, for $\left\vert E_{z}\right\vert >E_{\text{cr}}$, there
appear four bands crossing the gap at the K and K' points. When the Rashba
interaction is neglected ($\lambda _{\text{R}}=0$), each band contains
two-fold degenerate zero-energy states corresponding to up and down spins.
Even when $\lambda _{\text{R}}\neq 0$, the degeneracy remains unsolved at
the K (K') point, though it is slightly resolved away from these points. The
feature of the zero-energy states is highly contrasted with that in a
silicene nanoribbon, where they emerge at $k=\pi $ as in Fig.\ref%
{FigRibbonBand}(a). 
\begin{figure}[t]
\centerline{\includegraphics[width=0.45\textwidth]{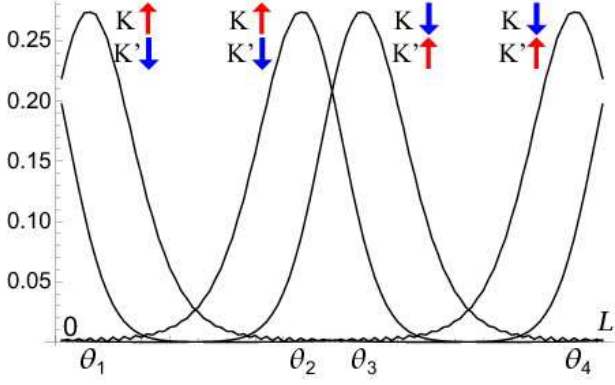}}
\caption{(Color online) The probability density of the zero modes in silicon
nanotube. There appear four peaks at $\protect\theta =\protect\theta _{j}$,
with $\sin \protect\theta _{j}=\pm E_{\text{cr}}/E$, as indicates the
emergence of four metallic states separating topological and band
insulators: See Fig.\protect\ref{FigHelicTube}. The holizontal axis is the
circumference ($0<x<L$) or the angle ($0<\protect\theta <2\protect\pi $). }
\label{FigJRWave}
\end{figure}

We show the probability density of these four zero modes in Fig.\ref%
{FigJRWave}, to which we have assigned the spin $s_{z}$ and the valley K$%
_{\mu }$ where they appear. We are able to make the assignment and determine
the direction of the current based on the effective Dirac theory: See a
sentence below (\ref{ThetaCriti}) and Fig.\ref{FigHelicTube}. Here we
explain them by using the gap formula (\ref{gap}) and the band structure in
Fig.\ref{FigTubeBand}(b).

It follows from the formula (\ref{gap}) that the electrons with spin $%
s_{z}=\pm 1$ are gapless at the K$_{\pm }$ point for $E_{z}>0$ and that the
electrons with spin $s_{z}=\mp 1$ are gapless at the K$_{\pm }$ point for $%
E_{z}<0$. Thus the peaks indexed by K$\uparrow $ appear in the upper half
and those indexed by K$\downarrow $ appear in the lower half of the
nanotube. Recall that $E_{z}$ is the electric field between the A and B
sublattices, which is opposite between the upper and lower halves of the
nanotube. Furthermore, the direction of the current can be determined by
examining the dispersion relation of the zero-energy mode in the band
structure [Fig.\ref{FigTubeBand}(b)]. We find the up-spin electrons and the
down-spin electrons propagate into the opposite directions in each peak.
Namely the current is helical, that is, it is a spin current, as illustrated
in Fig.\ref{FigHelicTube}. It is a characteristic feature that helical zero
modes appear along the edge of a topological insulator. We conclude that the
zero-energy metallic state separates the topological and band insulating
states in a silicon nanotube.

\section{Dirac theory}

In order to explore deeper physics of the helical zero modes, we analyze the
low-energy effective Hamiltonian derived from the tight binding model (\ref%
{BasicHamil}). It is described by the Dirac theory around the $K_{\eta }$
point as\cite{LiuPRB}%
\begin{equation}
H_{\eta }=\hbar v_{\text{F}}\left( k_{x}\tau _{x}-\eta k_{y}\tau _{y}\right)
+\eta \tau _{z}h_{11}+\ell E_{z}\tau _{z},  \label{DiracHamil}
\end{equation}%
with%
\begin{equation}
h_{11}=-\lambda _{\text{SO}}\sigma _{z}-a\lambda _{\text{R}}\left(
k_{y}\sigma _{x}-k_{x}\sigma _{y}\right) ,
\end{equation}%
where $\tau _{a}$ is the Pauli matrix of the sublattice pseudospin, $v_{%
\text{F}}=\frac{\sqrt{3}}{2}at=5.5\times 10^{5}$m/s is the Fermi velocity.

We have shown numerically that there emerge four zero-energy states inside
the bulk band gap as in Fig.\ref{FigTubeBand}(b). The low-energy Dirac
theory allows us to investigate analytically the properties of these helical
zero modes. In so doing we set $\lambda _{\text{R}}=0$ to simplify
calculations. This approximation is justified by the following reasons.
First all all, we have numerically checked that the band structure is rather
insensitive to $\lambda _{\text{R}}$ based on the tight-binding Hamiltonian (%
\ref{BasicHamil}). Second, $\lambda _{\text{R}}$ appears only in the
combination $(k_{x}\pm ik_{y})\lambda _{\text{R}}$ in the Hamiltonian (\ref%
{DiracHamil}), which vanishes exactly at the K$_{\pm }$ points. Third, the
critical electric field $E_{\text{cr}}$ is independent of $\lambda _{\text{R}%
}$ as in (\ref{CritiE}).

\begin{figure}[t]
\centerline{\includegraphics[width=0.45\textwidth]{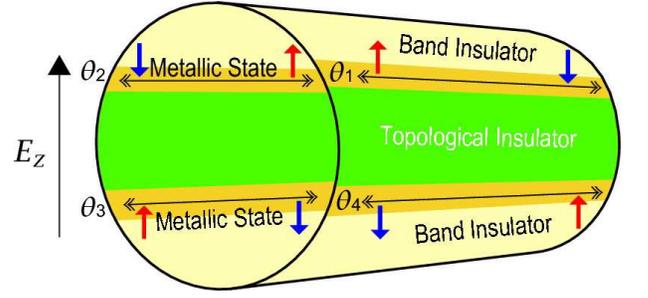}}
\caption{(Color online) An illustration of silicon nanotube under electric
field $E>E_{\text{cr}}$. There appear two topological insulator regions and
two band insulator regions. They are separated by metallic states made of
helical zero modes. A spin current flows in each metallic region as
indicated. For instance, up-spin (down-spin) electrons propagate into the
left (right) direction at $\protect\theta =\protect\theta _{1}$.}
\label{FigHelicTube}
\end{figure}

We take the $y$-axis parallel to the nanotube axis and the $x$-axis along
the circumference. We may set $k_{y}=$constant due to the translational
invariance along the $y$ axis. The momentum $k_{y}$ is a good quantum
number. Setting%
\begin{equation}
\Psi \left( x,y\right) =e^{ik_{y}y}\Phi \left( x\right) ,
\end{equation}%
we seek the zero-energy solution, where $\Psi \left( x,y\right) $ is a
four-component amplitude. The particle-hole symmetry guarantees the
existence of zero-energy solutions satisfying the relation $\phi _{B}\left(
x\right) =i\xi \phi _{A}\left( x\right) $ with $\xi =\pm 1$. Here, $\phi
_{A} $ is a two-component amplitude with the up spin and the down spin, $%
\phi _{A}=(\phi _{A}^{+},\phi _{A}^{-})^{t}$. Then the eigenvalue problem
yields%
\begin{equation}
H_{\eta }\phi _{A}(x)=E_{\eta \xi }\phi _{A}(x),
\end{equation}%
together with a linear dispersion relation%
\begin{equation}
E_{\eta \xi }=\eta \xi \hbar v_{\text{F}}k_{y}.  \label{DispeRelat}
\end{equation}%
The equation of motion for the component $\phi _{A}^{s_{z}}(x)$ reads%
\begin{equation}
\left( \xi \hbar v_{\text{F}}\partial _{x}+\eta s_{z}\lambda _{\text{SO}%
}-\ell E_{z}\left( x\right) \right) \phi _{A}^{s_{z}}(x)=0.  \label{DiracEqs}
\end{equation}%
We apply the uniform electric field $E$ perpendicular to the nanotube axis.

The effective field for electrons in a silicon nanotube is given by 
\begin{equation}
E_{z}(x)=E\sin \theta
\end{equation}%
with $\theta =2\pi x/L$. We take $E>E_{\text{cr}}$ with (\ref{CritiE}). We
solve the equation%
\begin{equation}
E_{z}(\theta )=\eta s_{z}E_{\text{cr}}.
\end{equation}%
Let us chose one solution and denote it as%
\begin{equation}
\theta _{\text{cr}}=\text{Arcsin}(E_{\text{cr}}/E).  \label{ThetaCriti}
\end{equation}%
We obtain two solutions $\theta _{1}=\theta _{\text{cr}}$ and $\theta
_{2}=\pi -\theta _{\text{cr}}$ for $\eta s_{z}=+1$, and two solutions $%
\theta _{3}=\pi +\theta _{\text{cr}}$ and $\theta _{4}=2\pi -\theta _{\text{%
cr}}$ for $\eta s_{z}=-1$. It implies that the zero modes at $\theta =\theta
_{1}$ contains up-spin electrons ($s_{z}=+1$) from the K valley ($\eta =+1$)
and down-spin electron ($s_{z}=-1$) from the K' valley ($\eta =-1$), and so
on, as illustrated in Fig.\ref{FigJRWave}. As we shall soon see, the sign $%
\xi $ is fixed at each $\theta _{j}$, as implies that up-spin and down spin
electrons propagate into the opposite directions according to the dispersion
relation (\ref{DispeRelat}). The current is helical, whose direction is
determined by the sign of $\xi $.

The equation of motion (\ref{DiracEqs}) is rewritten as 
\begin{equation}
\left( \frac{2\pi }{L}\xi \hbar v_{\text{F}}\partial _{\theta }-2\ell E\cos 
\frac{\theta +\theta _{j}}{2}\sin \frac{\theta -\theta _{j}}{2}\right) \phi
_{A}^{s_{z}}(\theta )=0.
\end{equation}%
To construct the solution near $\theta _{j}$, we approximate%
\begin{equation}
\cos \frac{\theta +\theta _{j}}{2}\simeq \cos \theta _{j}.
\end{equation}%
We can explicitly solve this as%
\begin{equation}
\phi _{A}^{s_{z}}\left( \theta \right) =C\exp \left[ -\xi \frac{2L\ell E}{%
\pi \hbar v_{\text{F}}}\cos \theta _{j}\cos \frac{\theta -\theta _{j}}{2}%
\right] ,
\end{equation}%
where $C$ is the normalization constant. It satisfies the periodic boundary
condition $\phi _{A}(\theta +\pi )=\phi _{A}(\theta )$ or $\phi
_{A}(x+L)=\phi _{A}(x)$. The sign $\xi $ is to be chosen so as to make the
peak of the wave function to appear at $\theta =\theta _{j}$, and we have%
\begin{equation}
\phi _{A}^{s_{z}}\left( \theta \right) =C\exp \left[ \frac{2L\ell E}{\pi
\hbar v_{\text{F}}}|\cos \theta _{j}|\cos \frac{\theta -\theta _{j}}{2}%
\right] ,
\end{equation}%
We obtain $\xi =-1$ at $\theta =\theta _{1}$ and $\theta =\theta _{4}$,
while $\xi =+1$ at $\theta =\theta _{2}$ and $\theta =\theta _{3}$. As we
have noticed in the sentence below (\ref{ThetaCriti}), the sign of $\xi $
determines the direction of the helical current at each $\theta _{j}$ based
on the dispersion relation (\ref{DispeRelat}), which we have illustrated in
Fig.\ref{FigHelicTube}. Furthermore, we have checked that the probability
density $|\phi _{A}^{s_{z}}\left( \theta \right) |^{2}$ agrees excellently
with the result obtained based on the tight-binding model in Fig.\ref%
{FigJRWave}.

\section{Conclusion}

We have uncovered a salient fact of silicon nanotube that it is a
topological insulator by studying its band structure in uniform electric
field $E_{z}$. When $E_{z}>E_{\text{cr}}$, there emerges four helical zero
modes propagating along the nanotube. They form metallic regions separating
topological and band insulators. We have constructed the wave function of
each zero mode based on the effective Dirac theory. Silicon nanotube may be
an ideal material to transport spin currents. It may be used as a quantum
wire in future spintronics. Our results are applicable also to germanium
nanotube.

\section{Acknowledgements}

I am very much grateful to N. Nagaosa for many fruitful discussions on the
subject. This work was supported in part by Grants-in-Aid for Scientific
Research from the Ministry of Education, Science, Sports and Culture No.
22740196.

\end{document}